\documentclass[reprint,amsmath,amssymb]{revtex4-1}
\usepackage[utf8x]{inputenc}
\usepackage{graphicx}% Include figure files
\usepackage{dcolumn}% Align table columns on decimal point
\usepackage{bm}% bold math
\usepackage{color}

\begin{document}

\preprint{APS/123-QED}

\title{Extended Creutz ladder with spin-orbit coupling: a
  one-dimensional analog of the Kane-Mele model}

\author{S. Gholizadeh}
\email{sina.gholizadeh@bilkent.edu.tr}
\affiliation{Department of Physics, Bilkent University, TR-06800 Bilkent, Ankara, Turkey}
\author{M. Yahyavi}
\email{m.yahyavi@bilkent.edu.tr}
\affiliation{Department of Physics, Bilkent University, TR-06800 Bilkent, Ankara, Turkey}
\author{B. Het\'enyi}
\email{hetenyi@fen.bilkent.edu.tr, hetenyi@phy.bme.hu}
\affiliation{Department of Physics, Bilkent University, TR-06800 Bilkent, Ankara, Turkey \\ and \\ MTA-BME Exotic Quantum Phases ``Momentum'' Research Group, Department of Physics, Budapest University of Technology and Economics, H-1111 Budapest, Hungary}

\begin{abstract}
  We construct a topological ladder model, one-dimensional, following
  the steps which lead to the Kane-Mele model in two dimensions.
  Starting with a Creutz ladder we modify it so that the gap closure
  points can occur at either $k = \pi / 2$ or $-\pi/2$.  We then
  couple two such models, one for each spin channel, in such a way
  that time-reversal invariance is restored.  We also add a Rashba
  spin-orbit coupling term.  The model falls in the CII symmetry
  class.  We derive the relevant $2\mathbb{Z}$ topological index,
  calculate the phase diagram and demonstrate the existence of edge
  states.  We also give the thermodynamic derivation of the quantum
  spin Hall conductance (St\v{r}eda-Widom).  Approximate
  implementation of this result indicates that this quantity is
  sensitive to the topological behavior of the model.
\end{abstract}


\maketitle

\section{Introduction}

Topological systems~\cite{Hasan10} are one of the most active current
research areas in condensed matter physics.  A crucial advance in this
field was the Haldane model~\cite{Haldane88} (HM), a hexagonal model
in which time-reversal symmetry and inversion symmetry are
simultaneously broken.  The model is engineered so that a gap can be
closed at either one of the Dirac points.  The gap closure occurs at a
phase line, which encloses a topological phase with finite Hall
conductance, whose sign depends on which gap is closed at the phase
line.  An extension of the HM, the Kane-Mele
model~\cite{Kane05a,Kane05b} (KMM), was another important step in the
development of topological insulators.  In this model two Haldane
models are taken, one for each spin channel, each one tuned so that
time-reversal symmetry is restored.  A Rashba coupling term, which
mixes different spins, is also added.  The KMM model exhibits
quantized quantum spin Hall (QSH) response, and sustains spin currents
at its edges.

Topological models in one
dimension~\cite{Su79,Rice82,Creutz94,Creutz99,Kitaev01,Li13,Li14,Wakatsuki14,Atherton16,Hetenyi18}
are also actively studied.  Of the many such models, most relevant to
our study is the Creutz model~\cite{Creutz94,Creutz99} which exhibits
a topological interference effect which can be probed when open
boundary conditions are applied (edge-states).  Recent
studies~\cite{Sticlet13b,Sticlet13,Viyuela14,Bermudez09} of this model
revealed several interesting phenomena.  The Uhlmann phase was
used~\cite{Viyuela14} as a measure of topological behavior at finite
temperature.  It was also shown~\cite{Bermudez09} that defect
production across a critical point obeys non-universal scaling
depending on the topological features.  We also emphasize that a
number of different one-dimensional topological
models~\cite{Li13,Hetenyi18} exhibit the same phase diagram as the HM.

Topological ladder models~\cite{Strinati17,Sun12} are one-dimensional
systems which, however, often exhibit effects usually associated with
two dimensions.  Strinati et al.~\cite{Strinati17} recently showed
that ladder models can support Laughlin-like states with chiral
current flowing along the legs of the ladder.  Since a ladder consists
of two legs separated by a finite distance, and enclosing a definite
area, it is possible to apply a magnetic field perpendicular to this
area and observe a quantum Hall response.  Another way to think about
this is to realize that to demonstrate the existence of chiral edge
currents, one needs a strip, which is also an effective
one-dimensional system, with a finite width (a ladder is a strip with
width of one, or a small number of, lattice constants).
Recently~\cite{Hetenyi18} we demonstrated that a ladder model can be
constructed to exhibit topological effects similar to the HM.  Our
interest here is whether it is possible to also construct a ladder in
the spirit of KMM.

In this paper we construct a ladder model, step-by-step, which can be
viewed as the one-dimensional analog of the KMM.  First, we modify the
original Creutz model so that gap closures are shifted in $k$-space,
breaking time-reversal invariance.  We then couple two such shifted
Creutz models, one for each spin channel, so that time-reversal
invariance is restored.  We also add a Rashba term to allow for the
mixing of spins.  We then derive a topological winding number for the
model, and calculate its phase diagram.  We also use the Widom
derivation of the QSH formula, which gives quantized response in the
topological region.  The possible experimental signature is spin
currents flowing along the legs of the ladder.

\begin{figure}[ht]
 \centering
 \includegraphics[width=\linewidth,keepaspectratio=true]{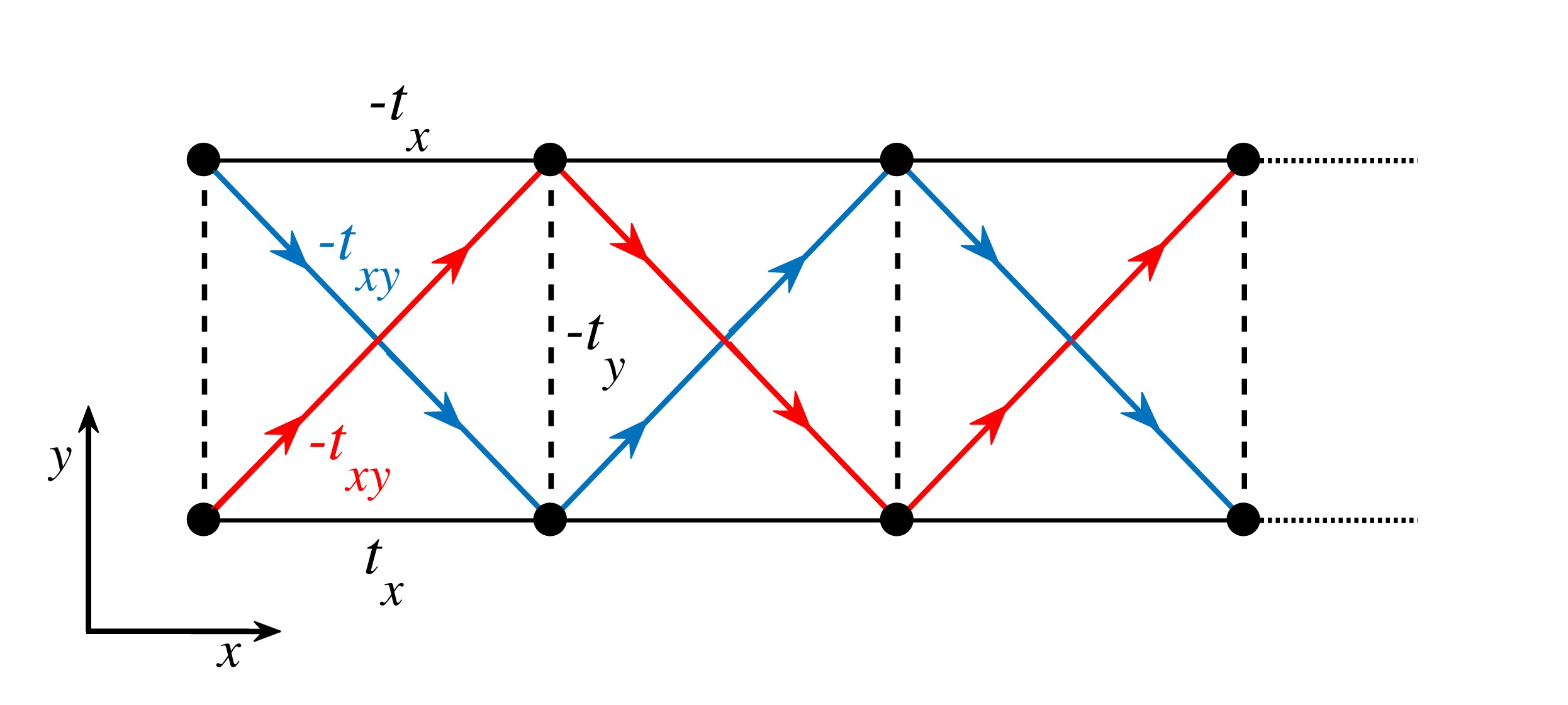}
 \caption{Graphic representation of our model.  $t_x$($t_y$) denotes
   hoppings along horizontal(vertical) bonds.  $t_{xy}$ denotes
   diagonal bonds.  We apply Peierls phases along the diagonal bonds
   along the directions indicated.  }
 \label{fig:ladder}
\end{figure}

\section{Models}

The Creutz model is a quasi-1D ladder model which exhibits a quantum
phase transition separating a trivial phase from a symmetry-protected
topological phase.  The topological phase is characterized by a
winding number, and if open boundary conditions are applied, localized
edge states are found.  Let $t_x$ denote hoppings along the legs,
$t_y$ the hoppings perpendicular to the legs, and $t_{xy}$ the
diagonal hoppings along unit cells.  In the original Creutz model a
magnetic field perpendicular to the plane of the system is applied,
resulting in Peierls phases along the legs of the ladder, pointing in
opposite directions on different legs of the ladder.  For a Peierls
phase of $\pi/2$ the resulting Hamiltonian is
\begin{equation}
  \label{eqn:H_C}
  H_C = - \sum_k [( 2 t_x \sin(k)) \hat{\sigma}_z + (t_y + 2 t_{xy} \cos(k) \hat{\sigma}_x)].
\end{equation}
Gap closure occurs at the points $k = 0,\pi$, depending on whether
$t_y = 2 t_{xy}$ or $t_y = -2 t_{xy}$.  Our first step is to set the
bonds on the upper(lower) leg to $t_x$($-t_x$) and introduce Peierls
phases of $\pi/2$ on the diagonal bonds as indicated in
Fig. \ref{fig:ladder}.  The Hamiltonian is now
\begin{equation}
  \label{eqn:H1}
  H_1 = - \sum_k [( 2 t_x \cos(k)) \hat{\sigma}_z + (t_y + 2 t_{xy}
    \cos(k + \Phi) \hat{\sigma}_x)].
\end{equation}
The first term alone corresponds to a band stucture
  with one-dimensional Dirac points at $k=\pm\pi/2$, which are
  time-reversal invariant pairs.  The second term opens gaps in
  general with masses of opposite signs at opposite Dirac points.
The phase diagram (determined by the gap closure condition) is the
same as that of the HM,
\begin{equation}
  \label{eqn:pdHM}
  \frac{t_y}{2 t_{xy}} = \pm \sin(\Phi).
\end{equation}
The sign in Eq. (\ref{eqn:pdHM}) determines which of the two gaps in
the Brillouin zone closes.

Given that the gap closures occur at time-reversal invariant points,
we can proceed to construct a one-dimensional analog of the
Kane-Mele model, by first introducing spin,
\begin{equation}
  \label{eqn:H2}
  H_2 = \sum_k d_1(k) \Gamma^{(1)}  + d_2(k) \Gamma^{(2)}  + d_{25}(k) \Gamma^{(25)},
\end{equation}
where we have used the following $\Gamma$-matrices, 
\begin{equation}
  \Gamma^{(a)} = \{ \sigma_x \otimes I_2, \sigma_z \otimes I_2, \sigma_y \otimes \sigma_x, \sigma_y \otimes \sigma_y, \sigma_y \otimes \sigma_z \},
\end{equation}
with $a=1,...,5$, and
\begin{equation}
  \Gamma^{(ab)} = \frac{1}{2i} [\Gamma^{(a)},\Gamma^{(b)}],
\end{equation}
$H_2$ can be viewed as the ``square'' of the $H_1$ Hamiltonian.  We
can now add the Rashba spin orbit coupling term resulting in
\begin{equation}
  \label{eqn:H}
  H = H_2 + d_3(k) \Gamma^{(3)} + d_{35}(k) \Gamma^{(35)}.
\end{equation}
The coefficients in Eqs. (\ref{eqn:H2}) and (\ref{eqn:H}) are given by
\begin{eqnarray}
  \label{eqn:dcoeffs}
d_1(k) = -t_y,d_2(k) = - 2 t_x \cos(k), d_3(k) = \lambda_R \\ \nonumber
 d_{25}(k) = 2 t_{xy} \sin(k),d_{35}(k) = 2 \lambda_R \sin(k).
\end{eqnarray}

\section{Symmetry analysis and topological indices}

Using the appropriate time-reversal, particle-hole and chiral symmetry
operators, one-dimensional models can be
placed~\cite{Altland97,Schnyder08} into topological classes.  For the
shifted Creutz model (Eq. (\ref{eqn:H_C})), the operator $T = i
\sigma_x K$($C = i \sigma_z K$, with $K$ denoting complex conjugation)
can be taken to be the time reversal (particle hole) operator, and the
time-reversal ($T^\dagger H(k) T = H(-k)$), partile-hole ($C^\dagger
H(k) C = -H(-k)$) and chiral symmetries ($S^\dagger H(k)S = - H(k)$,
where $S=TC$ is the chiral symmetry operator).  This implies that the
band-structure comes in pairs of $\pm\epsilon_k$.  $T^2 = C^2 = S^2 =
1$, placing the Creutz model in the BDI class.  For the shifted Creutz
model for $\Phi = \pi/2$ (Eq. (\ref{eqn:H1})) the time-reversal and
particle-hole symmetries are destroyed, but the chiral symmetry
remains ($S^\dagger H(k)S = -H(k)$), implying that the band-structure
again comes in pairs of $\pm\epsilon_k$.  The model falls in the
symmetry class AIII.

\begin{figure}[ht]
 \centering
 \includegraphics[width=\linewidth,keepaspectratio=true]{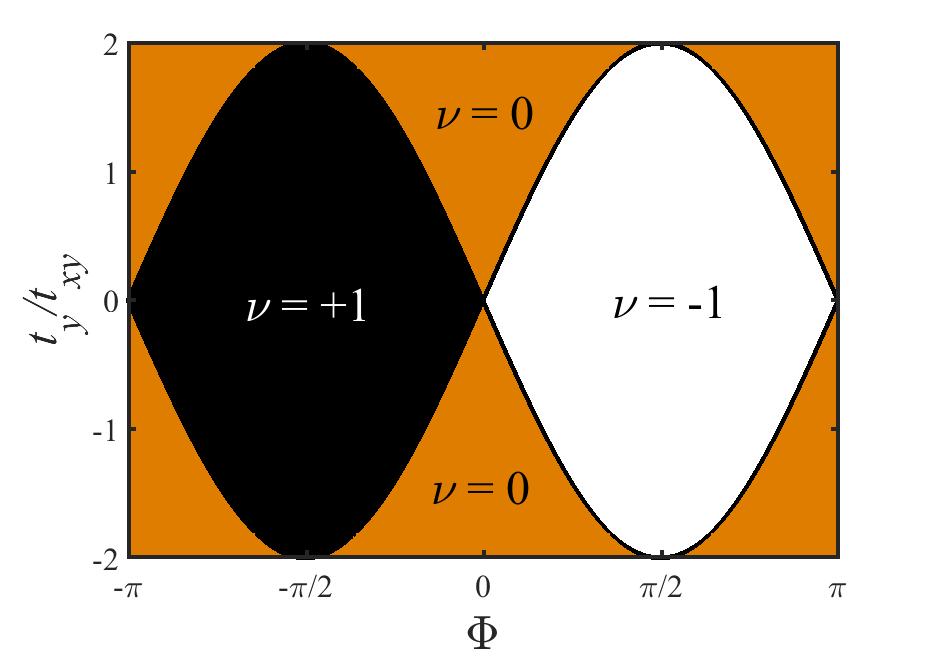}
 \caption{Phase diagram of the system where gap closure occurs. The
   numbers in the figures denote the topological winding number. }
 \label{fig:PD}
\end{figure}

For the spinful model we study, Eqs. (\ref{eqn:H2}) and (\ref{eqn:H})
the time-reversal and particle-hole operators take the form $T = i (
I_2 \otimes \sigma_y) K$, $C = i (\sigma_y \otimes \sigma_x) K$.  In
this case the square of the operators is $T^2 = C^2 = -1$, and $S^2 =
1$, placing these models in the CII symmetry class.  One can refine
the symmetry characterization further by also considering the
reflection operator~\cite{Chiu13,Chiu16}, which sends $k$ to $-k$
without altering the spin.  This operator is $R = ( I_2 \otimes
\sigma_x)$, which anti-commutes with $T$, but commutes with $C$.  In
terms of mirror symmetry class~\cite{Chiu13,Chiu16} the model falls in
class $C$, with a topological index of $2 \mathbb{Z} $.

For the a chiral symmetric Hamiltonian (Eq. (\ref{eqn:H1})) we apply a
unitary transformation~\cite{Ryu10}, constructed from spinors of spin
in the $y$ direction,
\begin{eqnarray}
 U = \frac{1}{\sqrt{2}} \begin{pmatrix}
1 & 1 \\
i  & -i
\end{pmatrix} ,
\label{eqn:transform-matrix_U}
\end{eqnarray}
to our Hamiltonian.  This leaves us with the off-diagonal form,
\begin{eqnarray}
  \mathcal{H}_{T} = U^{\dagger} \mathcal{H} U  = \hspace{2cm} \\ \nonumber = 2t_x \cos(k) \sigma_x + [t_y - 2t_{xy}\sin(k)] \sigma_y =
  \begin{pmatrix}
 0 & q \\
 q^{\dagger} & 0
\end{pmatrix}.
\end{eqnarray}
where $q=2t_x \cos(k) -i[ t_y - 2t_{xy}\sin(k)]$.
The winding number density is given by,
\begin{eqnarray}
w(k) = i q^{-1}(k)\partial_{k}q(k),
\end{eqnarray}
from which the winding number can be obtained by integrating across
the full Brillouin zone after setting $t_x = t_{xy} = 1$, resulting in
\begin{eqnarray}
\mathcal{W} = \int_{-\pi}^{\pi} \frac{dk}{2\pi} w(k),
\end{eqnarray}
which can be turned into a contour integral around the unit circle via
$z = e^{ik}$.  If the point $(0,i t_y/2)$ is within the ellipse
defined by $(2 t_x \cos(k),2 t_{xy} \sin(k))$ with $0\leq k < 2 \pi$,
the winding number is minus one.  Otherwise it is zero.

We proceed to extend this result to Eq. (\ref{eqn:H2}).  In this case
we have 4-by-4 block-diagonal Hamiltonian,
\begin{eqnarray}
& \mathcal{H}(k) =
 2 t_x \cos(k)  \sigma_0 \otimes \sigma_z -2t_{xy}\sin(k)  \sigma_z \otimes \sigma_x  + t_y  \sigma_0 \otimes \sigma_x  \nonumber \\
& = \begin{pmatrix}
h^{\uparrow} & 0 \\
0 & h^{\downarrow}
\end{pmatrix}
\end{eqnarray}
where 
$$h^{\uparrow,\downarrow} = \begin{pmatrix}
2t_x \cos(k) & t_y\mp2t_{xy} \sin(k) \\
t_y\mp2t_{xy}\sin(k) & -2t_x \cos(k)
\end{pmatrix}$$  
 After transforming Hamiltonian under $(\sigma_0 \otimes U)$, where $U$ is given by Eq. (\ref{eqn:transform-matrix_U}),
\begin{eqnarray}
\mathcal{H}_{T}(k) = (\sigma_0 \otimes U)^{\dagger} \; \mathcal{H}(k) \; (\sigma_0 \otimes U)
= \begin{pmatrix}
h^{\uparrow}_{T} & 0 \\
0 & h^{\downarrow}_{T}
\end{pmatrix}
\end{eqnarray}
where $h^{\uparrow}_{T} = \begin{pmatrix}
0 & q_1 \\
q_1^{\dagger} & 0
\end{pmatrix}$ and $h^{\downarrow}_{T} = \begin{pmatrix}
0 & q_2 \\ 
q_2^{\dagger} & 0
\end{pmatrix}$ while
\begin{eqnarray}
q_1 = 2t_{x}\cos(k)-i(t_y - 2t_{xy}\sin(k))  \nonumber \\
q_2= 2t_{x}\cos(k)-i(t_y + 2t_{xy}\sin(k)) .
\end{eqnarray}
Notice that our $4\times4$ Hamiltonian is simply two independent
Creutz models.  The overall winding number will be the sum of the
winding number of each Creutz model, the two possible values therefore
are minus two or zero, depending on whether the point $z = -it_y / 2$
falls inside or outside the ellipse defined by the Brillouin zone,
respectively.

The fundamental group corresponding to topological index of the
Hamiltonian of each spin channel is $\mathbb{Z}$.  The space of
$\mathcal{H}_T$ is decomposed into direct sum of subspaces of
$h^{\uparrow}$ and $h^{\downarrow}$:
\begin{eqnarray}
\mathcal{H}_T = h^{\uparrow} \oplus h^{\downarrow},
\end{eqnarray}
hence, the fundamental group representation of topological index can
be written as sum of fundamental groups of two subspaces,
\begin{eqnarray}
2\mathbb{Z} = \mathbb{Z} + \mathbb{Z}
\end{eqnarray}
which is consistent with symmetry analysis outcome.

\section{St\v{r}eda-Widom formula for quantum spin Hall systems}

In the case of the QH effect, a very
useful~\cite{Yilmaz15,Hetenyi18} formula was derived by
St\v{r}eda~\cite{Streda82} via quantum transport equations, and also
by Widom~\cite{Widom82} via thermodynamic Maxwell relations.  The
generalization to the QSH effect, similar to the
St\v{r}eda approach, was done by Yang and Chang~\cite{Yang06}.  Here
we attempt to derive this via Widom's thermodynamic considerations.

As a starting point, we take the view that a topological insulator
consists of two magnets of opposite polarization for each spin.  We
also invoke a spin-dependent magnetic field, and a corresponding
spin-dependent vector potential, ${\bf B}_\sigma$ and ${\bf
  A}_\sigma$, respectively.  Such a procedure was recently applied by
Dyrdal {\it et al.}~\cite{Dyrdal16} to calculate the properties of a
two-dimensional electron gas with Rashba spin-orbit coupling.  Under
the first assumption the spin current can be written as
\begin{equation}
  {\bf J}_{SH} = c \nabla \times \left[ {\bf M}_\uparrow - {\bf M}_\downarrow \right].
\end{equation}
We can derive the electric field from the chemical potential as ${\bf
  E} = \nabla (\mu/e)$, we can write the spin current as 
\begin{equation}
  {\bf J}_{SH} = (ec) {\bf E} \times
\frac{\partial}{\partial \mu}
  \left[ {\bf M}_\uparrow - {\bf M}_\downarrow \right].
\end{equation}
We can apply the Maxwell relation and arrive at
\begin{equation}
  {\bf J}_{SH} = {\bf E} \times
  \left[
    \frac{\partial (n e c)}{\partial{{\bf B}_\uparrow}}
    - 
    \frac{\partial (n e c)}{\partial{{\bf B}_\downarrow}},
    \right].
\end{equation}
resulting in a QSH conductivity of 
\begin{equation}
  \label{eqn:sigmaSH}
  \sigma_{SH} = ec
  \left[
    \frac{\partial n }{\partial{B_\uparrow}}
    - 
    \frac{\partial n }{\partial{B_\downarrow}}
    \right]_\mu,
\end{equation}
where we took the magnetic fields for both spins to be pointing
perpendicular to the plane (justifying the neglect of tensor
notation).  We can rewrite this expression in terms of particle number
and magnetic flux as
\begin{equation}
  \sigma_{SH} = 
  \left[
    \frac{\partial \nu }{\partial{\Phi_\uparrow}}
    - 
    \frac{\partial \nu }{\partial{\Phi_\downarrow}}
    \right]_\mu.
\end{equation}
This expression points to a definite procedure to calculate
$\sigma_{SH}$; calculate the Fermi level in the absence of flux, then
introduce a spin-dependent flux quantum, and count the number of
particles which cross the Fermi level.  In our approximate
implementation, we use equal and opposite flux for the different spin
channels on the $t_{xy}$ bonds.  Following Dyrdal {\it et
  al.}~\cite{Dyrdal16} we neglect the effect of the spin-dependent
vector potentials on the Rashba spin-orbit coupling term.

\section{Results}

\begin{figure}[ht]
 \centering
 \includegraphics[width=\linewidth,keepaspectratio=true]{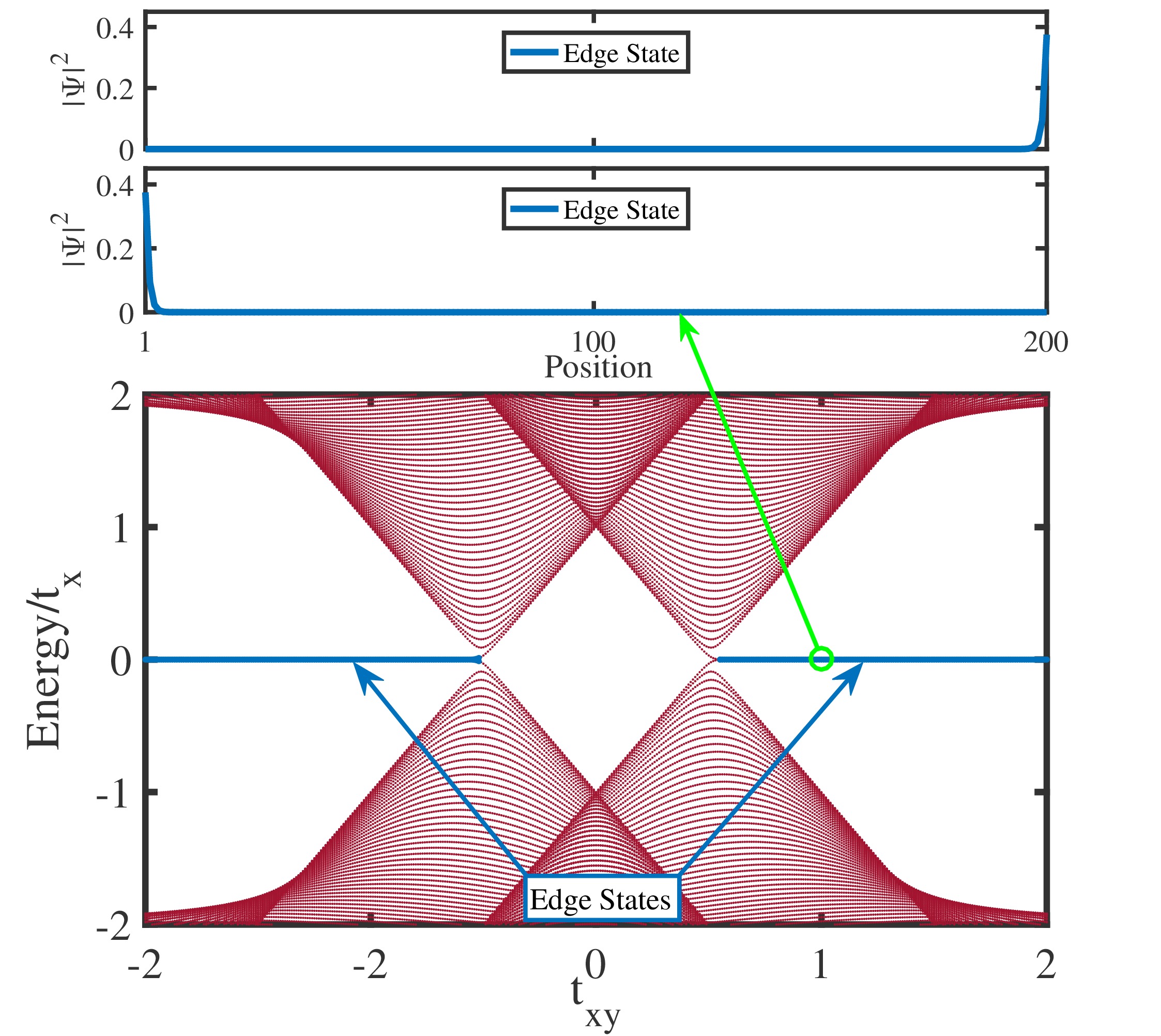}
 \caption{Energy spectrum of the shifted Creutz model with $200$
   sites, open boundary conditions as a function of $t_{xy}$.  The
   parameters are $t_x=1$ and $t_y =1$.  The blue lines indicate
   states which are not present when periodic boundary conditions are
   applied.  The square magnitude of these states are shown in the
   upper panels.  They are localized near the edges of the chain.}
 \label{fig:edge-state-spinless}
\end{figure}

The gap in the band structure of the shifted Creutz model closes at $k
= -\pi/2$ and $\pi/2$ depending on whether $t_y = 2 t_{xy}$ or $t_y =
- 2 t_{xy}$.  When the boundary conditions are open edge states are
found as shown in the shifted Creutz model
(Fig. \ref{fig:edge-state-spinless}).  The combination of two shifted
Creutz models, one for each spin, restore time reversal invariance
with gap closures at $k = \pm \frac{\pi}{2}$.  Obviously, this system
will also exhibit edge states.

\begin{figure}[ht]
 \centering
 \includegraphics[width=\linewidth,keepaspectratio=true]{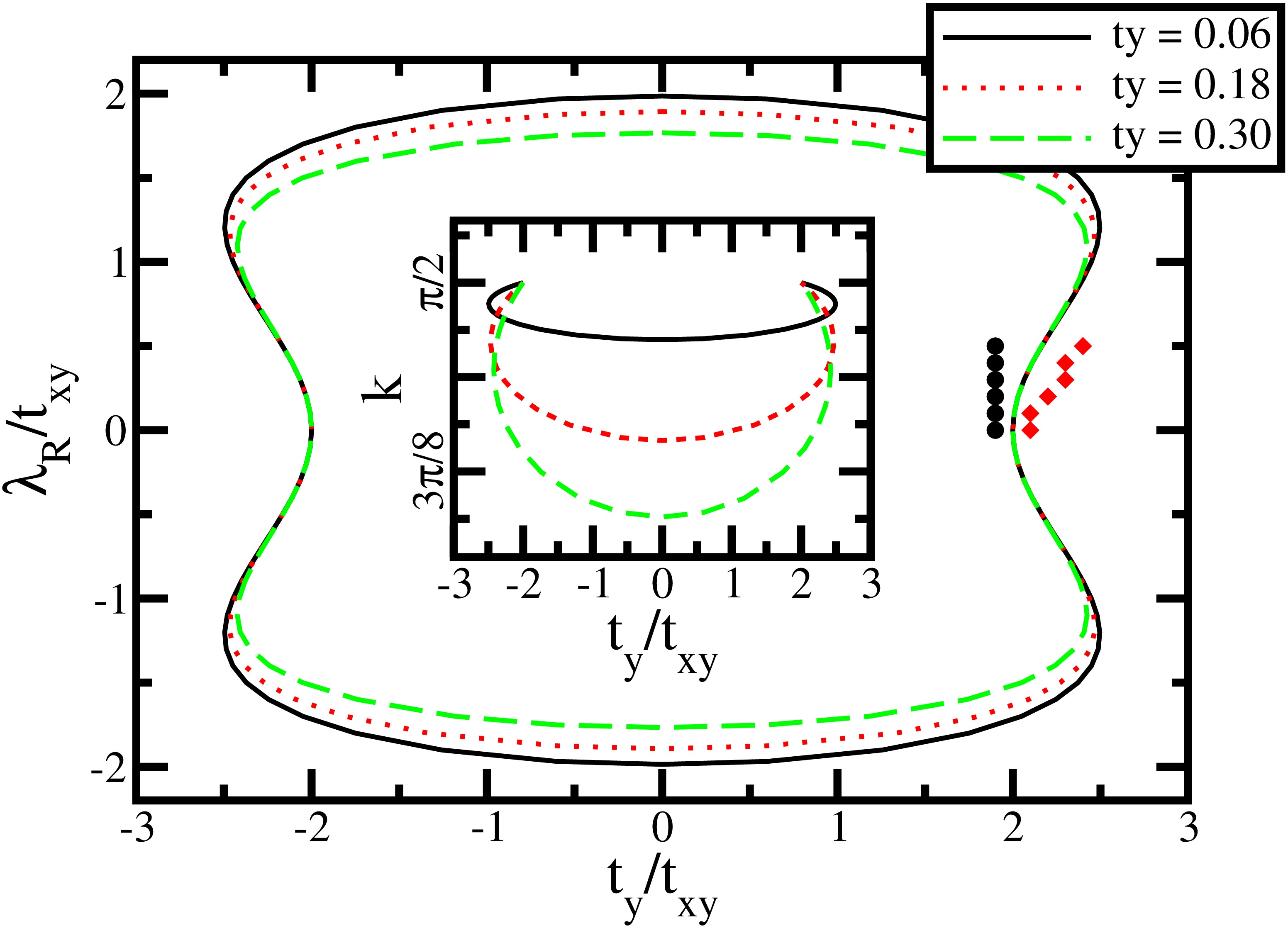}
 \caption{Main figure: phase diagram for systems with $t_x=1, t_{xy} =
   0.06,0.12,0.18$ in the $t_y$ vs. $\lambda_R$ plane.  The inset
   indicates the $k$ vector at which gap closure occurs as a function
   of $t_y/t_{xy}$.  The black filled circles and red filled diamonds
   on the left side of the phase diagram indicate systems for which we
   have tested our St\v{r}eda-Widom formula.  For the black filled
   circles we found a quantized quantum spin Hall response, while we
   found no response for the red diamonds.}
 \label{fig:pd_k}
\end{figure}

Turning on the Rashba coupling term gives rise to a phase diagram
shown in Fig. \ref{fig:pd_k} for three cases.  The plots are based on
a calculation in which $t_x=1$, $t_{xy} = 0.03,0.18$ and $0.30$.  The
phase diagram in the $\lambda_R/t_{xy}$ vs. $t_y/t_{xy}$ is shown for
these three cases.  The topological phase is the one which includes
the origin, outside of this region the phase is trivial.  The lines
indicate where gap closure occurs.  Along the phase boundary the
system becomes an ideal conductor with a finite Drude weight.  The
inset shows the absolute value of the $k$ points at which the gap
closure occurs as a function of $t_y/t_{xy}$.  

We also studied the quantized transport properties of the models,
based on the approximate implementation of the result in
Eq. (\ref{eqn:sigmaSH}).  For no Rashba coupling we find that the
trivial phase exhibits no $\sigma_{SH}$ response, in other words, upon
turning on the spin-dependent flux on the diagonal bonds leads to no
change in the number of particles below the Fermi level.  In the
topological phase, the flux decreases the number of particles under
the Fermi level by two.  For small values of the Rashba coupling
$\lambda_R \approx 0.5 t_{xy}$ we find the same.  In
Fig. \ref{fig:pd_k} we indicate the points at which we made
calculations (black filled circles and red filled diamonds).  At
larger values of $\lambda_R$ our approximations appear to break down.
However, we emphasize that the topological region is adiabatically
connected to the $\lambda_R = 0$ region and is therefore the same
quantum phase (also characterized by the $2\mathbb{Z}$ winding number
derived above).  

\section{Conclusion}

In conclusion we have assembled a one-dimensional ladder analog of the
Kane-Mele model, step by step, first by ``shifting'' the Creutz model
in the Brillouin zone, then introducing spin and spin-orbit coupling.
Our model falls in the CII symmetry class.  We also derived a formula
for the quantum spin Hall response and made an approximate
implementation.  For small values of the Rashba coupling, where our
approximation is expected to be valid, we find a quantized spin Hall
response in the topological phase indicating that QSH
  currents flowing along the legs of the ladder are a unique feature
  exhibited by our model.

The experimental realization of our model can most likely be done with
cold atoms in optical lattices.  Standard one-dimensional
models~\cite{Bloch08} already have some history in this setting, but
even more complex ones, such as multi-orbital ladder model with
topologically non-trivial behavior can be realized~\cite{Sun12}.
There are several interesting routes, for example, it is possible to
construct~\cite{Strinati17} optical lattices with cold atoms in which
the atomic states play the role of spatial indices, a technique known
as synthetic dimension.  A more difficult aspect is the presence of
spin-orbit couplings.  In two dimensions this was only done
recently~\cite{Grusdt17}, via a combination of microwave driving and
lattice shaking.  A key development in this experiment is that the
different spin-orbit couplings can be varied independently, therefore
Kane-Mele like models can be built.

\end{document}